\documentclass[twocolumn,amssymb,nobibnotes,aps,prbstab, reprint,letterpaper,pre/printnumbers,amsmath,
superscriptaddress,longbibliography,
showpacs,floatfix]{revtex4-1}
\usepackage{graphicx,xcolor}
\usepackage{dcolumn}
\usepackage{bm}
\usepackage{graphics}
\usepackage{float}
\usepackage{epsfig}
\usepackage{multirow}
\usepackage{amsmath}
\usepackage{multirow}
\usepackage{float}
\usepackage{caption}
\usepackage[justification=raggedright,singlelinecheck=false]{caption}
\usepackage{tabularx}
\usepackage{bm}        

\usepackage{mathrsfs}
\usepackage{tabularx}
\usepackage{tabulary}
\usepackage{gensymb}
\usepackage{enumitem}
\usepackage[dvipsnames]{xcolor}
\usepackage[pass,paperwidth=8.5in,paperheight=11in]{geometry}
\usepackage[linktocpage=true]{hyperref}
\usepackage{hyperref}
\usepackage{comment}
\usepackage{soul}
\usepackage{notes2bib}
\hypersetup{
  pdfnewwindow=true, 
  colorlinks=true,
  linkcolor=blue, 
  anchorcolor=blue,
  citecolor=blue, 
  filecolor=blue,
  menucolor=blue, 
  urlcolor=blue}

\usepackage{gensymb}
\usepackage{array}
\usepackage[normalem]{ulem}
\usepackage{xcolor}
\usepackage{orcidlink}

\begin{document}


\title{Pressure-Driven Structural Phase Competition and Functional Response in Layered LiInP$_2$S$_6$}

\author{Xiaochi Xie}
\affiliation{Department of Mechanical Engineering, University of Rochester, Rochester, New York 14627, USA}
\affiliation{Department of Mechanical Engineering, Stanford University, Stanford, CA 94305, USA}

\author{Pegah Mohammadi}
\affiliation{Department of Mechanical Engineering, University of Rochester, Rochester, New York 14627, USA}

\author{Sobhit Singh}
\email{s.singh@rochester.edu}
\affiliation{Department of Mechanical Engineering, University of Rochester, Rochester, New York 14627, USA}
\affiliation{Materials Science Program, University of Rochester, Rochester, New York 14627, USA}

\begin{abstract}
Understanding how hydrostatic pressure modifies interlayer interactions and competing ionic configurations is essential for controlling the emergent functional properties of layered quantum materials.
Here, using first-principles density-functional theory calculations, we investigate the pressure-dependent structural, mechanical, electronic, and optical properties of three competing LiInP$_2$S$_6$ polymorphs: the monoclinic $C2/c$ phase and the trigonal $P\bar{3}1c$ phase in both in-layer and in-gap configurations. 
Our results reveal a pressure-induced structural phase transition from the monoclinic ground-state $C2/c$ phase to a trigonal $P\bar{3}1c$ in-layer phase at $\sim$0.38\,GPa, driven by enhanced interlayer coupling and anisotropic lattice compression. 
In contrast, the trigonal $P\bar{3}1c$ in-gap phase remains energetically unfavorable due to its stronger interlayer ionic interactions and reduced compressibility. 
All phases remain mechanically stable under compression (0--26\,GPa) and exhibit enhanced mechanical rigidity, elastic wave velocities, and Debye temperatures with increasing pressure.
Remarkably, the electronic and optical properties within each phase remain highly robust under pressure, with only moderate changes in the band gap and optical absorption edge (UV-Visible range) under pressure; however, substantial modifications emerge across the pressure-induced structural phase transition. 
These findings establish LiInP$_2$S$_6$ as a pressure-sensitive ionic-vdW material in which subtle changes in interlayer interactions govern structural stability and functional properties.

\end{abstract}

\maketitle


\section{\label{sec:level1}Introduction}

Layered van der Waals (vdW) materials have emerged as a fertile platform for discovering unique electronic, ferroic, and ionic phenomena arising from reduced dimensionality and weak interlayer coupling ~\cite{liu2019van, Wang2023NatMater,fei2018ferroelectric,Geim2013Nature, Novoselov2016Science, xiao2018intrinsic, Ajayan2016PhysicsToday, chaturvedi2020universal}. 
In these systems, even subtle perturbations such as strain, electric fields, intercalation, or hydrostatic pressure can dramatically reconstruct the energy landscape by modifying the delicate balance between intralayer bonding and interlayer interactions \cite{brehm2020tunable, valasek1921piezo, gui2025stacking, shah_pressure-tuned_2026, liu2023stress, rao2021pressure, Grzechnik1998SSC}. 
As a result, layered materials frequently host competing structural phases, tunable electronic states, and emergent collective responses that are absent in conventional three-dimensional crystals.


Among layered compounds, transition-metal thiophosphates with chemical formula $M_x$P$_2$S$_6$ ($M$ = metal ion) have attracted immense attention because they combine ionic degrees of freedom with vdW-layered architectures~\cite{du2020conversion, neumayer2022ionic, sun2019strain,Mohammadi2025PRB, xu2024emerging, song2024anisotropic, studenyak2003optical, Kim2023ACSANM, kleemann2011magnetic, susner2017metal, kuhn2014synthesis}.
These materials exhibit an exceptional range of functionalities including ferroelectricity \cite{Liu2016NatCommun, dziaugys2010dipolar, jiang2024dual, morozovska2024ferri, he2023unconventional, Belianinov2015NanoLett, maisonneuve1997ferrielectric, sharma2019room}, ferroionic switching \cite{Mohammadi2025PRB, seleznev2023cyclic, neumayer2025competing, li2024realization, liang2025configurable, rao2021pressure, Grzechnik1998SSC, simon1994paraelectric, bourdon199931p}, giant negative electrostriction \cite{you2019origin, neumayer2019giant, Neumayer2020AEM}, pressure-induced structural transitions ~\cite{shah_pressure-tuned_2026, Mohammadi2025PRB, rao2021pressure, Grzechnik1998SSC, sun2019strain, morozovska2021stress, morozovska2023strain}, ionic conductivity \cite{zhang2021anisotropic, maisonneuve1997ionic}, and unconventional polarization dynamics~\cite{seleznev2023cyclic, reimers2018van, Mohammadi2025PRB}.
In particular, CuInP$_2$S$_6$ has emerged as a prototypical vdW ferroionic system where the competition between ionic motion and layered bonding gives rise to rich pressure-dependent phenomena~\cite{shah_pressure-tuned_2026, neal2022vibrational, seleznev2023cyclic, Mohammadi2025PRB, liu2019van, morozovska2023strain, liu2022excited, simon1994paraelectric, he2023unconventional, rao2021pressure, Grzechnik1998SSC}.
Similarly, AgInP$_2$S$_6$ represents another member of this layered family, exhibiting ferroelectricity, ferro- or antiferromagnetic behavior, as well as piezoelectric properties, along with promising optical, and thermoelectric characteristics~\cite{dziaugys2010dipolar, wang2016second, babuka2018structural, zhang2014electrically, peeters1997circularly}. 
Together, these studies have unequivocally established layered thiophosphates as an important materials family for investigating how external perturbations manipulate coupled structural and electronic degrees of freedom.


Very recently, LiInP$_2$S$_6$ has emerged as a particularly intriguing member of the layered-thiophosphate family. 
Liang \textit{et al.}~demonstrated spontaneous water intercalation within the vdW gap of LiInP$_2$S$_6$, leading to enhanced ionic conductivity and highlighting its potential for superionic conduction and inorganic iontronic devices~\cite{liang2024anomalous}. Further investigations by Qian \textit{et al.} revealed that this material exhibits promising optoelectronic properties~\cite{doi:10.1021/acs.chemmater.4c01876}, while Bai \textit{et al.} proposed LiInP$_2$S$_6$/XTe$_2$ (X = Mo, W) heterostructures as candidates for next-generation optoelectronic applications~\cite{doi:10.1021/acs.jpcc.5c01999}. 


Experimental synthesis identified a trigonal $P\bar{3}1c$ phase~\cite{Liang2024SmartMat}, while recent first-principles studies revealed several nearly degenerate polymorphs distinguished by distinct Li configurations~\cite{mohammadi_first-principles_2026}. 
Notably, the monoclinic $C2/c$ phase and the trigonal $P\bar{3}1c$ in-layer configuration differ in energy by only a few meV per unit cell~\cite{mohammadi_first-principles_2026}. 
Such near degeneracy strongly suggests that LiInP$_2$S$_6$ resides near a structural instability where modest external perturbations and/or thermal effects can reorganize the underlying potential-energy landscape.

Hydrostatic pressure provides a unique and clean route to probe this competition because it directly modifies the vdW gap, interlayer orbital hybridization, and local ionic coordination without introducing chemical disorder in metal thiophosphates~\cite{fernandes1991effect, kusmartseva2009pressure, souza2009pressure, telford2023designing, hu2023two, zhou2024sliding, wang2025chemical}. 
For example, in CuInP$_2$S$_6$, hydrostatic pressure has been reported to drive a monoclinic-to-trigonal phase transition, alter ferroelectric polarization, and even induce metallization~\cite{shah_pressure-tuned_2026, Mohammadi2025PRB, zhou2024sliding, bu2022enhanced, rao2021pressure, Rao2023JCP, luo2024pressure}.        
Despite extensive pressure studies in CuInP$_2$S$_6$, the pressure response of LiInP$_2$S$_6$ remains yet unexplored.

\begin{figure}[!htb]
\centering
\includegraphics[width=\columnwidth]{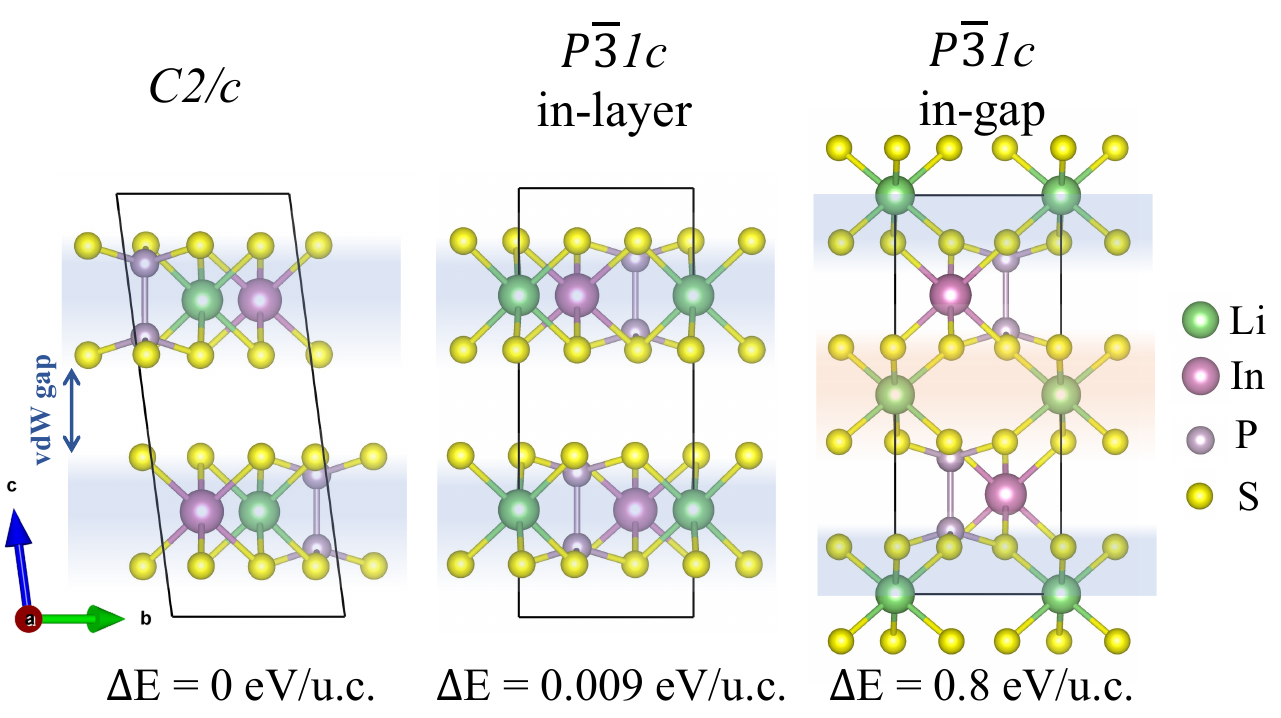}
\caption{Crystal structures of layered LiInP$_2$S$_6$. The material exhibits a monoclinic $C2/c$ phase and two trigonal $P\overline{3}1c$ configurations depending on the position of the Li atom: within the layer (in-layer) or in the vdW gap (in-gap). The relative energies of the structures are shown with respect to the ground-state $C2/c$ phase at ambient conditions. }
\label{fig:structure}
\end{figure}

Here, using first-principles density-functional theory (DFT) calculations, we investigate the pressure-dependent structural, mechanical, electronic, and optical properties of the three known polymorphs of LiInP$_2$S$_6$ shown in Fig.~\ref{fig:structure}~\cite{mohammadi_first-principles_2026}. 
Our results reveal that hydrostatic pressure stabilizes the trigonal in-layer phase ($P\bar{3}1c$) through enhanced interlayer coupling and anisotropic lattice compression, driving a structural phase transition from the monoclinic ground-state $C2/c$ structure to the trigonal $P\bar{3}1c$ in-layer phase at 0.38\,GPa. 
All the three studied phases remain elastically and mechanically stable under the studied pressure range (0 -- 26\,GPa). 
We further uncover a strong interplay between vdW-gap reduction, elastic stiffening, 
and electronic structure evolution under compression.
It is worth noting that the band gaps of the LiInP$_2$S$_6$ polymorphs remain remarkably robust against hydrostatic pressure. 
The maximum change in the predicted bandgap is about 13\% at 10\,GPa hydrostatic pressure. 
Overall, our findings establish LiInP$_2$S$_6$ as an important model system for studying pressure-controlled ionic-vdW coupling in layered materials.


\section{\label{sec:method}Methodology}

First-principles density functional theory (DFT) were performed within the framework of projector augmented-wave (PAW) method as implemented in the Vienna \textit{Ab initio} Simulation Package (VASP)~\cite{PhysRevB.54.11169, KRESSE199615, PhysRevB.59.1758, PhysRevB.50.17953}. In the PAW pseudopotentials, the valence electronic configurations considered were Li ($2s^{1}$), In ($5s^{2}5p^{1}$), P ($3s^{2}3p^{3}$), and S ($3s^{2}3p^{4}$). 
The exchange-correlation effects were treated within the generalized gradient approximation (GGA) using the Perdew--Burke--Ernzerhof for solids (PBEsol) functional~\cite{PhysRevLett.100.136406}. 
Structural relaxations were carried out until the residual Hellmann--Feynman forces on each atom were smaller than $10^{-3}$~eV/\AA, and the total electronic self-consistent energy was converged to $10^{-7}$~eV.

Since the studied material is a layered vdW system, long-range dispersion interactions were taken into account using the DFT-D3 method~\cite{10.1063/1.3382344, https://doi.org/10.1002/jcc.21759}. 
Brillouin-zone integrations were performed using a $\Gamma$-centered $8 \times 8 \times 4$ $k$-point mesh. 
Hydrostatic pressure was applied through full variable-cell structural relaxation. The relative phase stability was evaluated using enthalpy differences.

The elastic constants $C_{ij}$ were carefully converged with respect to the $k$-point sampling. A detailed analysis of the elastic properties, including the evaluation of longitudinal, transverse, and average elastic wave velocities, as well as the Debye temperature, was carried out using the \textsc{MechElastic} Python package~\cite{singh_mechelastic_2021, singh_PRB2018}. 
The electronic band structures and projected density of states were analyzed using the \textsc{PyProcar} package~\cite{herath2020pyprocar, lang2024expanding}.
The frequency-dependent optical properties were computed within the independent-particle approximation using the frequency-dependent dielectric tensor implemented in VASP~\cite{PhysRevB.73.045112}. 
Post-processing of the dielectric function, refractive index, and absorption coefficient was performed using the \textsc{VASPKIT} package~\cite{wang2021vaspkit}.

\begin{figure}[!htbp]
\centering
\includegraphics[width=0.45\textwidth]{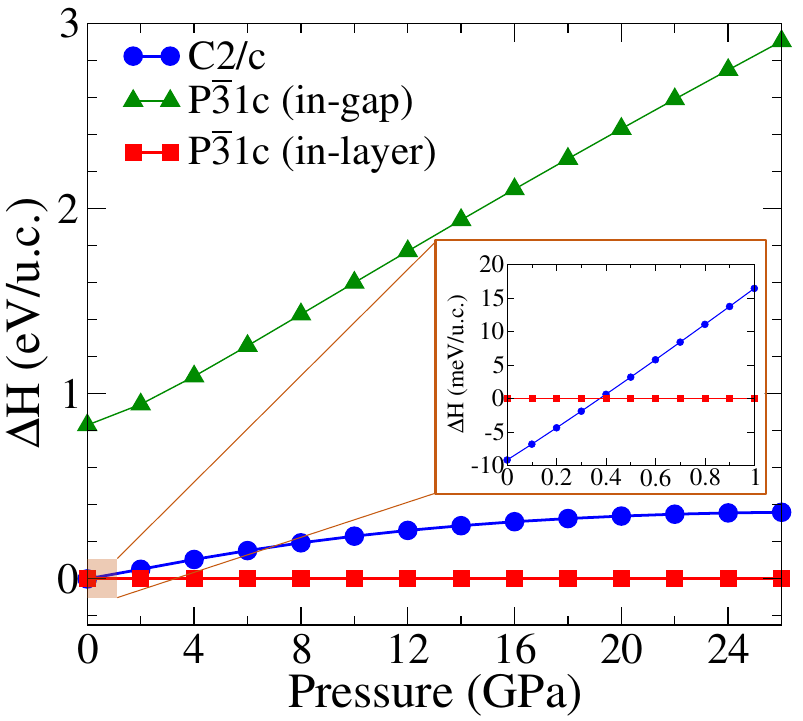} 
\caption{Enthalpy difference ($\Delta H = H - H_{P\overline{3}1c\,(\text{in-layer})}$) versus hydrostatic pressure phase diagram.}
\label{fig:enthalpy}
\end{figure}

\section{\label{sec:results}Results and Discussion}

\subsection{\label{sec:struct}Pressure-driven phase competition and anisotropic compression}

For vdW LiInP$_2$S$_6$ crystals, three stable phases have been previously reported~\cite{mohammadi_first-principles_2026}: a monoclinic $C2/c$ phase (space group no.~15) and two trigonal $P\bar{3}1c$ phases (space group no.~163). 
In the trigonal $P\bar{3}1c$ configuration, two distinct Li arrangements are possible: (i) the in-layer configuration, where Li atoms are located within each LiInP$_2$S$_6$ monolayer, and (ii) the in-gap configuration, where Li atoms occupy the vdW gap between adjacent layers, as shown in Fig.~\ref{fig:structure}. 
At ambient pressure, the $C2/c$ phase is energetically favored, although the energy difference relative to the trigonal in-layer configuration is approximately 9\,meV per unit cell.


\begin{figure*}[!htbp]
\centering
\includegraphics[width=1\textwidth]{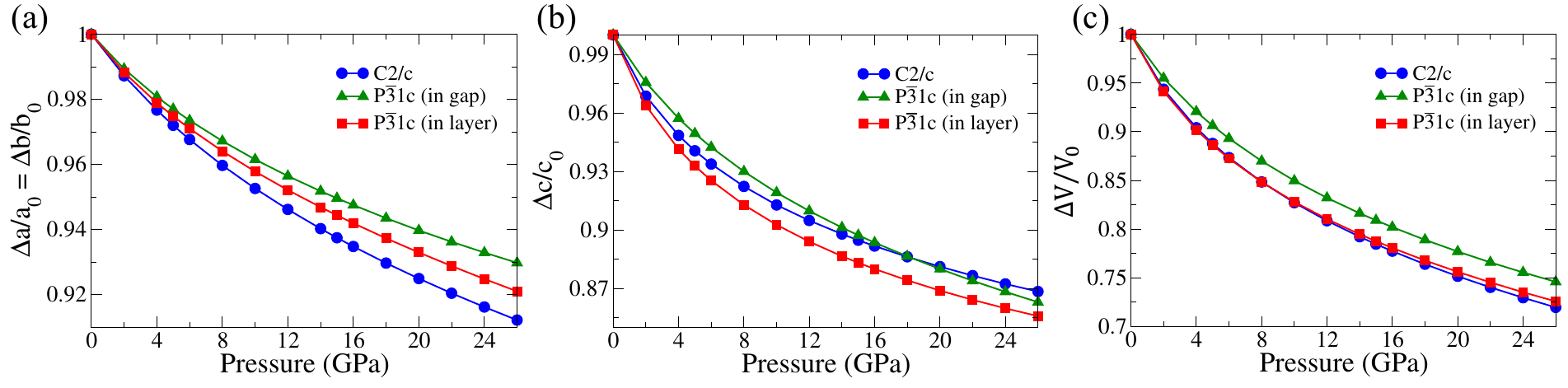} 
\caption{Pressure dependence of the normalized lattice parameters relative to their values at 0 GPa for the studied phases under increasing hydrostatic pressure. (a) In-plane lattice parameters $a=b$, (b) Out-of-plane lattice parameter $c$, and (c) Normalized unit-cell volume $V$ as a function of pressure.}
\label{fig:lattice}
\end{figure*}

Figure~\ref{fig:enthalpy} presents the relative enthalpy differences ($\Delta H$) of the studied phases with respect to the $P\bar{3}1c$ (in-layer) structure.
The results reveal that pressure drives a phase transition from the monoclinic $C2/c$ phase to the trigonal $P\bar{3}1c$ (in-layer) phase at 0.38\,GPa. 
This trigonal phase remains enthalpically stable up to the maximum applied pressure of 26\,GPa. 
In contrast, the in-gap configuration becomes progressively less favorable with increasing pressure.
Thus, compression favors Li coordination within the layer rather than within the vdW gap.

This behavior reflects the competition between ionic coordination and interlayer compression. In the trigonal in-layer phase, pressure reduces the vdW gap and strengthens interlayer coupling without introducing significant steric repulsion around the Li atoms. 
In contrast, the in-gap structure resists compression because the Li ions occupy the interlayer vdW region, leading to a mixed ionic-covalent-type of bonding with the neighboring sulfur atoms. To further determine the ionic contribution of the interlayer Li--S interaction, the percentage ionic character (\%IC) was estimated using Pauling’s electronegativity~\cite{matsunaga2003pauling}: \%IC = $\left(
1-e^{-0.25(\Delta \chi)^2}
\right)\times100$, 
where $\Delta \chi$ represents the absolute difference in the Pauling electronegativities of the two bonded atoms.
%
By considering electronegativity values of 0.98 for Li and 2.58 for S, the ionic character of the Li--S interaction is estimated to be approximately 47\%. As a result, the in-gap phase exhibits reduced compressibility and remains energetically unfavorable throughout the investigated pressure range.


The evolution of the lattice parameters and unit-cell volume of the studied phases under pressure is examined in Fig~\ref{fig:lattice}.
With increasing pressure, all three phases exhibit anisotropic lattice contraction, consistent with the directional bonding nature of layered metal thiophosphates such as CuInP$_2$S$_6$~\cite{Mohammadi2025PRB, zhou2024sliding}. 
As expected, the $P\bar{3}1c$ (in-gap phase), shows a stronger resistance to compression. 
Consequently, its lattice parameters, particularly along the out-of-plane direction, decrease more gradually than those of the other candidate phases, which are dominated by weaker vdW interactions.
These results highlight the pressure-dependent structural response and motivate us for a detailed investigation of the mechanical properties. 
Therefore, in the next section, the elastic and mechanical properties are systematically analyzed.


\begin{table*}[htbp]
\centering
\small
\renewcommand{\arraystretch}{1.2}
\setlength{\tabcolsep}{4pt}
\caption{Elastic stiffness constants $C_{ij}$ (in GPa) for the studied LiInP$_2$S$_6$ phases}
\begin{tabular}{l c c c c c c c c c c c c c c c}
\hline\hline
\textbf{Phase} & \textbf{Pressure (GPa)} 
& $C_{11}$ & $C_{22}$ & $C_{33}$ & $C_{44}$ & $C_{55}$ & $C_{66}$ 
& $C_{12}$ & $C_{13}$ & $C_{14}$ & $C_{15}$ & $C_{23}$ & $C_{25}$ & $C_{35}$ & $C_{46}$ \\
\hline

\multirow{2}{*}{$C2/c$} 
 & 0  & 95.4 & 95.9 & 42.4 & 9.8 & 9.6 & 35.5 & 24.5 & 7.0 & -0.1 & 1.2 & 7.1 & -1.2 & -0.2 & -1.1 \\
\cline{2-16}
 & 10 & 150.2 & 148.7 & 132.4 & 32.6 & 31.4 & 49.9 & 50.3 & 35.1 & -0.1 & 5.6 & 34.6 & -5.2 & -0.3 & -4.9 \\
\hline

 \multirow{2}{*}{$P\bar{3}1c$ (in-layer)} 
 & 0  & 100.0 & 100.1 & 37.1 & 9.2 & 9.3 & 37.3 & 25.4 & 6.4 & 0.0 & -3.3 & 6.4 & 3.2 & -0.1 & 3.1 \\
\cline{2-16}
 & 10 & 166.3 & 166.3 & 120.3 & 39.5 & 39.6 & 58.9 & 48.0 & 42.8 & 0.0 & -17.6 & 42.8 & 17.5 & -0.1 & 17.4 \\
\hline

\multirow{2}{*}{$P\bar{3}1c$ (in-gap)} 
 & 0  & 92.3 & 92.3 & 55.2 & 16.6 & 16.6 & 29.6 & 32.7 & 17.1 & 0.0 & -1.1 & 17.0 & 1.1 & -0.1 & 1.0 \\
\cline{2-16}
 & 10 & 153.1 & 153.0 & 109.3 & 41.2 & 41.2 & 44.5 & 63.6 & 52.8 & 0.0 & -2.2 & 52.7 & 2.2 & 0.0 & 2.1 \\
\hline\hline

\end{tabular}
\label{tab:elastic_const}
\end{table*}

\begin{table*}[htbp]
\centering
\small
\renewcommand{\arraystretch}{1.2}
\setlength{\tabcolsep}{4pt}
\caption{Elastic moduli and derived mechanical properties for the studied LiInP$_2$S$_6$ phases}
\label{tab:elastic_prop}
\begin{tabular}{l c c c c c c c c c c}
\hline\hline
\textbf{Phase} & \textbf{Pressure (GPa)} & \textbf{$K$ (GPa)} & \textbf{$G$ (GPa)} & \textbf{$E$ (GPa)} & \textbf{$\nu$} & \textbf{$K/G$} & \textbf{$v_l$ (m/s)} & \textbf{$v_t$ (m/s)} & \textbf{$v_m$ (m/s)} & \textbf{$\Theta_D$ (K)} \\
\hline

\multirow{2}{*}{$C2/c$}
 & 0  & 31.4 & 20.2 & 49.8 & 0.24 & 1.55 & 4327.8 & 2546.3 & 2822.0 & 309.2 \\
\cline{2-11}
 & 10 & 74.2 & 42.1 & 106.1 & 0.26 & 1.76 & 5883.9 & 3342.6 & 3716.1 & 433.8 \\
\hline

\multirow{2}{*}{$P\bar{3}1c$ in-layer}
 & 0  & 30.6 & 19.9 & 49.0 & 0.24 & 1.54 & 4263.3 & 2517.2 & 2789.0 & 306.5 \\
\cline{2-11}
 & 10 & 79.0 & 45.4 & 114.4 & 0.26 & 1.74 & 6065.5 & 3461.2 & 3846.5 & 450.2 \\
\hline
 
\multirow{2}{*}{$P\bar{3}1c$ in-gap}
 & 0  & 39.6 & 23.0 & 57.9 & 0.26 & 1.72 & 4766.6 & 2727.2 & 3030.1 & 331.4 \\
\cline{2-11}
 & 10 & 82.3 & 41.5 & 106.6 & 0.28 & 1.98 & 6145.7 & 3375.3 & 3762.7 & 434.5 \\
\hline\hline

\end{tabular}
\end{table*}

\subsection{\label{sec:mech}Pressure-driven mechanical and elastic properties}

To evaluate the mechanical and elastic behavior of the studied phases, key quantities such as Young's modulus ($E$), Poisson's ratio ($\nu$), bulk modulus ($K$), and shear modulus ($G$) are calculated. The Voigt--Reuss--Hill (VRH) averaging scheme~\cite{hill1952elastic} is employed to evaluate the bulk and shear moduli, from which Young’s modulus and Poisson’s ratio are subsequently derived. These properties provide essential insight into the response of the material under external pressure and its evolution with compression.

According to the calculated elastic stiffness coefficients $C_{ij}$ (Table~\ref{tab:elastic_const}), all diagonal components of $C_{ij}$ matrix are positive and the full tensor satisfies the required symmetry relations for all three studied phases at both ambient and high pressure. 
Moreover, the calculated elastic tensors $C_{ij}$ for all three phases satisfy the Born-Huang mechanical stability criteria, as implemented in the {\sc MechElastic} package~\cite{singh_mechelastic_2021}, both at ambient conditions and under applied pressure. Our results verify that the studied structures remain mechanically stable under pressure. 

\begin{figure}[!htb]
\centering
\includegraphics[width=0.5\textwidth]{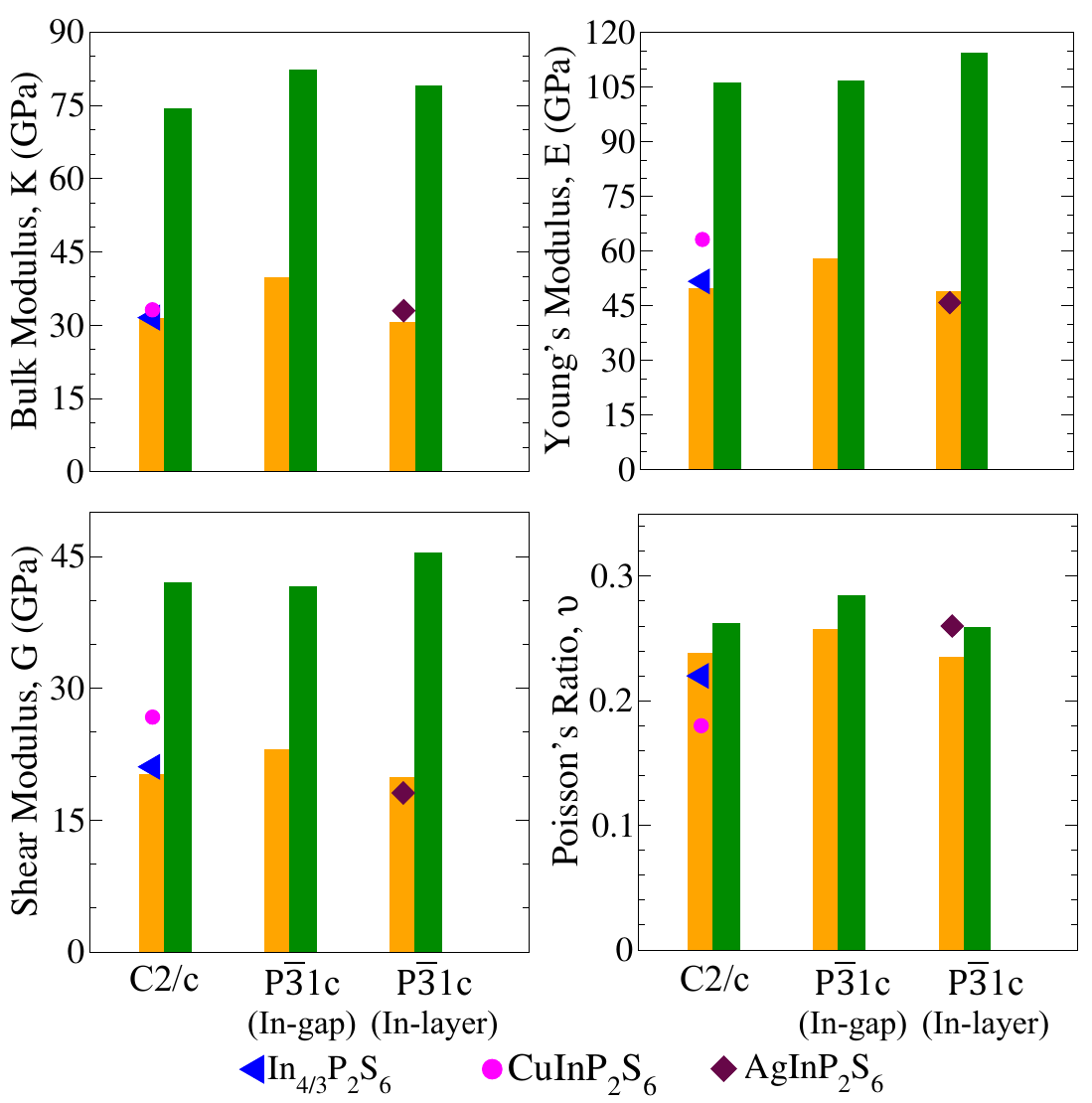} 
\caption{Elastic constants of LiInP$_2$S$_6$ for the $C2/c$, $P\overline{3}1c$ (in-gap), and $P\overline{3}1c$ (in-layer) phases. Orange-colored bars correspond to values calculated at 0 GPa, while the green bars indicate those obtained at 10\,GPa. The symbols $\blacktriangleleft$, $\bullet$, and $\blacklozenge$ correspond to similar metal thiophosphate compounds reported in Ref.~\cite{babuka2018structural, babuka2026electronic}.}
\label{fig:mechanical}
\end{figure}

Notably, at ambient pressure, the in-gap trigonal phase exhibits the largest elastic moduli $K, G, E$, and $\nu$, as shown in Fig.~\ref{fig:mechanical} and tabulated in Table~\ref{tab:elastic_prop}, indicating intrinsically higher rigidity under compression. 
This behavior originates from the stronger effective interlayer coupling associated with Li occupation inside the vdW region, which suppresses layer sliding and lattice deformation. 
All three phases become mechanically stiffer under compression, while the energetically favorable $P\bar{3}1c$ (in-layer) phase exhibits approximately 10\% higher enhancement in the shear and Young’s moduli compared to the $C2/c$ and $P\bar{3}1c$ (in-gap) phases at 10 GPa.

We further compare the elastic properties of LiInP$_2$S$_6$ phases with related metal thiophosphate compounds reported in the literature ~\cite{babuka2026electronic,babuka2018structural}. In particular, In$_{4/3}$P$_2$S$_6$ and CuInP$_2$S$_6$, both in the monoclinic phase~\cite{babuka2026electronic}, as well as AgInP$_2$S$_6$ in the trigonal $P\bar{3}1c$ (in-layer) phase~\cite{babuka2018structural}, have been previously studied using DFT (GGA/PBE-D) methods. Their mechanical properties are included in Fig.~\ref{fig:mechanical} for comparison. It is observed that the elastic moduli of In$_{4/3}$P$_2$S$_6$ show closer agreement with those of LiInP$_2$S$_6$ than with CuInP$_2$S$_6$. This behavior can be attributed to the similarity in the valence electronic configurations of In ($5s^2 5p^1$) and Li ($2s^1$), whereas Cu ($3d^{10}4s^1$) possesses a filled $d$ shell, 
leading to a stronger hybridization and a more pronounced covalent bonding (in the context of percentage ionic-covalent character of bonds) with surrounding S atoms, and consequently higher stiffness. Furthermore, AgInP$_2$S$_6$, which has the same trigonal symmetry as the $P\bar{3}1c$ (in-layer) phase, exhibits similar mechanical behavior as compared to LiInP$_2$S$_6$.


Next, we analyze the ductile versus brittle behaviour the studied phases by calculating the $K/G$ ratio, where values greater than 1.7 indicate ductile behavior~\cite{singh_mechelastic_2021, kojima2024poisson}. 
At ambient pressure, the vdW phases ($C2/c$ and $P\bar{3}1c$ in-layer) exhibit $K/G < 1.7$, indicating brittle behavior. However, with increasing pressure, their $K/G$ ratios exceed 1.7, signaling a pressure-induced transition toward ductility. In contrast, the $P\bar{3}1c$ in-gap phase maintains $K/G > 1.7$ across the entire pressure range, indicating intrinsically ductile behavior due to its stronger interlayer bonding, which is further enhanced under compression.

\begin{figure*}[!htb]
\centering
\includegraphics[width=1\textwidth]{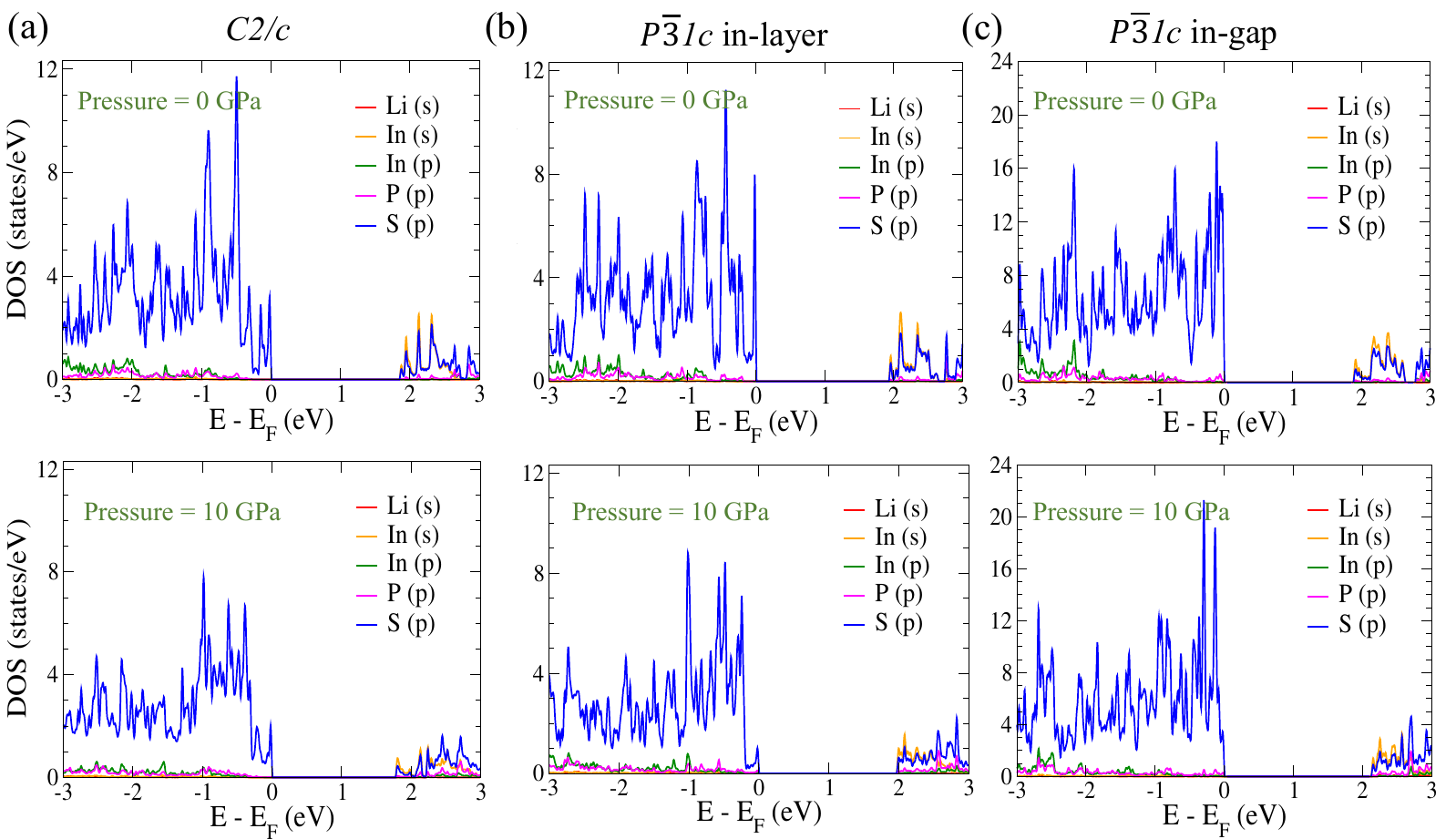} 
\caption{Calculated orbital-resolved density of states (DOS) for LiInP$_2$S$_6$: (a) $C2/c$ phase, (b) $P\overline{3}1c$ (in-layer), and (c) $P\overline{3}1c$ (in-gap) phase at 0 and 10 GPa.}
\label{fig:DOS}
\end{figure*}

Using the {\sc MechElastic} package, we further evaluate the longitudinal ($v_l$), transverse ($v_t)$, and average ($v_m$) elastic wave velocities, along with the Debye temperature ($\Theta_D$) at both ambient conditions and under compression. The obtained data are provided in Table~\ref{tab:elastic_prop}. 
At ambient pressure, the trigonal $P\bar{3}1c$ in-gap  phase exhibits higher wave velocities than the monoclinic $C2/c$ and trigonal $P\bar{3}1c$ in-layer  phases, indicating faster propagation of elastic waves, which is consistent with its stronger in-gap bonding and higher stiffness. 
Under compression, all three phases show enhanced wave velocities due to reduced interatomic distances and strengthened bonding interactions, reflecting increased lattice rigidity.

\begin{figure*}[!htb]
\centering
\includegraphics[width=1\textwidth]{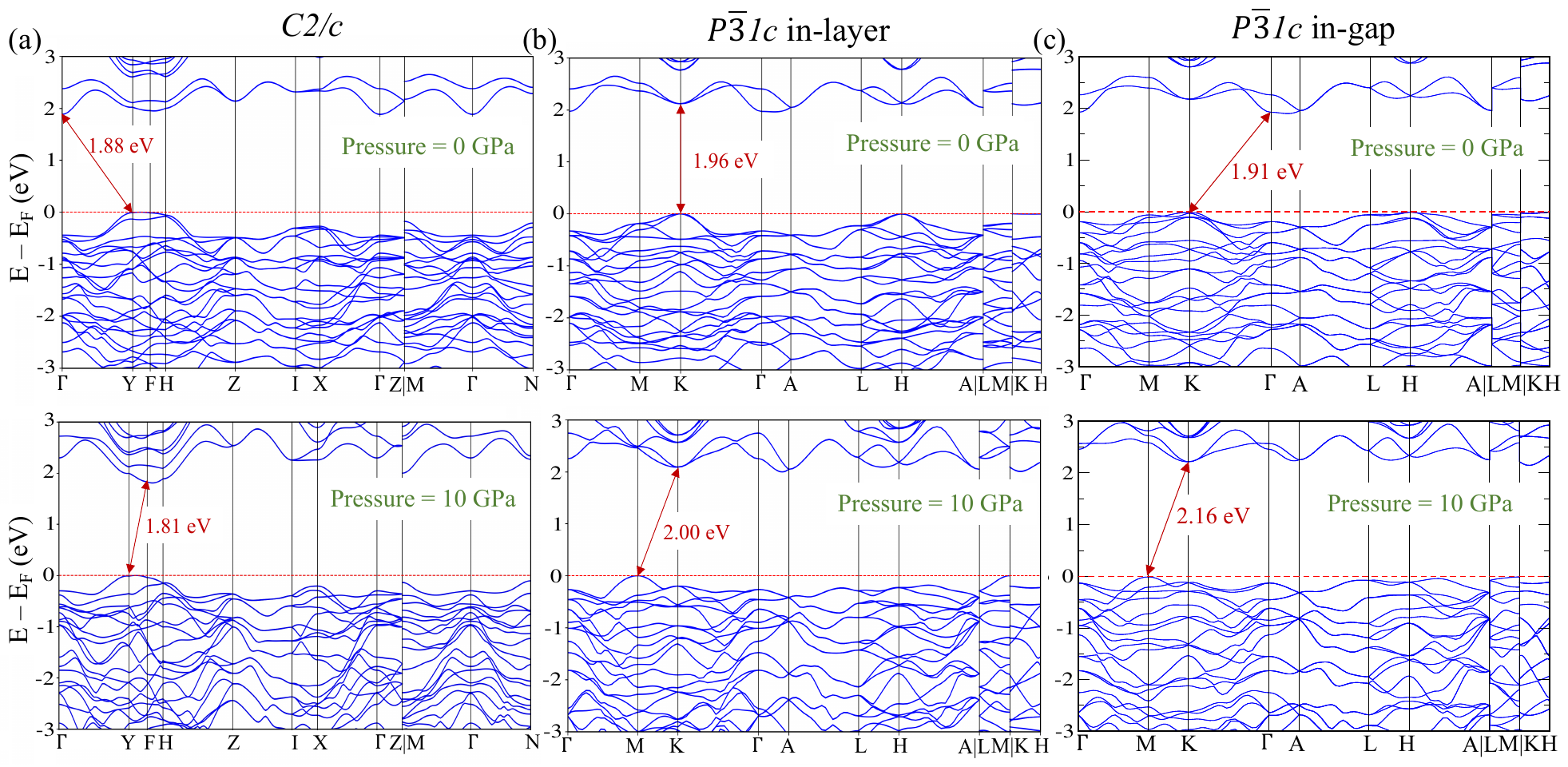} 
\caption{Electronic band structures of LiInP$_2$S$_6$: (a) $C2/c$ phase, (b) $P\overline{3}1c$ (in-layer), and (c) $P\overline{3}1c$ (in-gap) phase at 0 and 10 GPa. The dashed horizontal line indicates the Fermi level ($E_F$).}
\label{fig:bands}
\end{figure*}

A similar trend is observed for the Debye temperature $\Theta_D$, which is a key thermodynamic parameter for better understanding quantum effects on the lattice vibrations, specific heat, and thermal conductivity,
particularly under applied pressure~\cite{singh_mechelastic_2021}. 
Since $\Theta_D$ depends on the average elastic wave velocity and material density, it systematically increases under pressure. 

Overall, these results reveal a pronounced pressure-driven enhancement of lattice rigidity and dynamical stability, governed by the evolution of interlayer interactions in LiInP$_2$S$_6$ polymorphs.

\subsection{\label{sec:mech}Electronic structure evolution and interlayer hybridization under compression} 

In this section, we examine the electronic properties of the $C2/c$ and $P\bar{3}1c$ (in-layer and in-gap) phases by analyzing both the atom-projected density of states (DOS) and the electronic band structures near the Fermi level ($E_F$), calculated along the high-symmetry $k$-path in the Brillouin zone.
The DOS is calculated at both ambient pressure and under applied pressure (10\,GPa). 
As shown in Fig.~\ref{fig:DOS}, at ambient conditions, the valence band in all phases is predominantly composed of S-$p$ orbitals, while the conduction band arises from hybridization between In-$s$ and S-$p$ orbitals. Under applied pressure, although the crystals experience compression, the dominant ionic contribution between In and S atoms preserves the overall electronic character of the phases. Consequently, the orbital contributions and hybridization behavior remain largely unchanged, indicating that the electronic bonding characteristics are maintained under pressure.

The calculated electronic band structures using the PBEsol functional are also investigated at both ambient and high pressure for the three polymorphs (see Fig.~\ref{fig:bands}). After confirming that the band gap values are in good agreement with previous studies~\cite{doi:10.1021/acs.jpcc.5c01999, mohammadi_first-principles_2026}, we analyze their evolution under applied pressure. For the monoclinic $C2/c$ phase, the indirect band gap decreases slightly from 1.88~eV to 1.81~eV with increasing pressure, corresponding to a reduction of approximately 3\%. On the other hand, for the trigonal phases, including both the in-layer and in-gap configurations, the band gap marginally increases under compression. In particular, the band gap of the $P\bar{3}1c$ (in-layer) phase increases by approximately 2\%, while the $P\bar{3}1c$ (in-gap) phase exhibits a larger enhancement of about 13\%, indicating a stronger pressure response of the electronic structure in the in-gap configuration. Interestingly, for the $P\bar{3}1c$ in-layer phase, which becomes energetically favorable under pressure, the band gap acquires an indirect nature along the $M-K$ k-path.


The different pressure trends highlight the sensitivity of the electronic structure to the local Li coordination environment. In particular, the electronic response of the in-layer phase is governed not simply by bond shortening, but by the competition between intralayer and interlayer orbital interactions.



\begin{figure*}[!htb]
\centering
\includegraphics[width=1\textwidth]{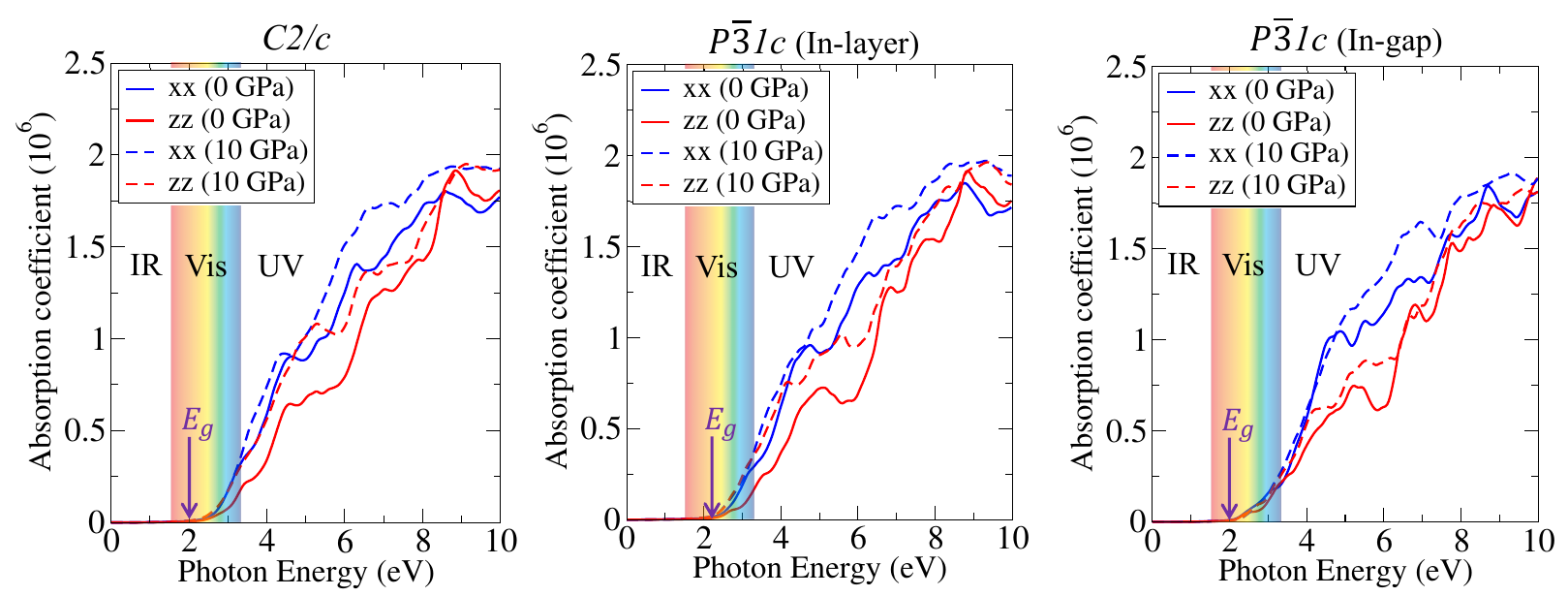}
\caption{Calculated optical absorption coefficients of LiInP$_2$S$_6$ for (a) $C2/c$, (b) $P\bar{3}1c$ (in-layer), and (c) $P\bar{3}1c$ (in-gap) phases at 0 and 10\,GPa. For each phase, the in-plane ($xx$) and out-of-plane ($zz$) components are shown to illustrate the optical anisotropy and its evolution under pressure within the semilocal independent-particle approximation. Solid lines correspond to 0\,GPa, while dashed lines represent 10\,GPa.}
\label{fig:absorption}
\end{figure*}

\subsection{\label{sec:optics}Pressure-driven optical response}

Motivated by the electronic structure results, which place the band gaps of LiInP$_2$S$_6$ polymorphs near the UV--Vis range, we investigate their optical properties under applied pressure. The frequency-dependent complex dielectric function,
\begin{equation}
\varepsilon(\omega)=\varepsilon_1(\omega)+i\varepsilon_2(\omega),
\end{equation}
is calculated within the independent-particle approximation (IPA) for all three phases at ambient pressure and 10~GPa (see Fig.~\ref{fig:dielectric}). 
Based on the Kramers--Kronig relations~\cite{de1926theory, gajdovs2006linear}, $\varepsilon_1(\omega)$ and $\varepsilon_2(\omega)$ enable the evaluation of key optical quantities, including the absorption coefficient $\alpha(\omega)$, energy-loss function $L(\omega)$, extinction coefficient $k(\omega)$, and refractive index $n(\omega)$, using the standard expressions in literature ~\cite{solyom2009optical, de1926theory}.

The optical absorption coefficient, which reflects the penetration of light into the material, is shown in Fig.~\ref{fig:absorption} for the $C2/c$, $P\bar{3}1c$ (in-layer), and $P\bar{3}1c$ (in-gap) phases.  Within the PBEsol framework, the absorption onset for all phases lies in the visible range, as indicated by the purple arrows in each absorption spectrum shown in Fig.~\ref{fig:absorption}. 
The calculated absorption spectra exhibit pronounced anisotropy between the in-plane and out-of-plane components, which is expected for  layered vdW structures. At ambient conditions, all phases show optical absorption edges in the visible energy range (at the PBEsol level). 
As reported in previous studies~\cite{mohammadi_first-principles_2026, doi:10.1021/acs.jpcc.5c01999}, the hybrid-functional (HSE) calculations predict band gaps that are approximately 1.2\,eV larger than those obtained using PBEsol, suggesting that the actual absorption edge may extend further into the UV--Vis region. 

Our results indicate that the optical response of each LiInP$_2$S$_6$ phase remains relatively robust under pressure. The band gap changes by only approximately 3\% and 2\% for the vdW $C2/c$ and $P\bar{3}1c$ (in-layer) phases, respectively, and by about 13\% for the bulk-like $P\bar{3}1c$ (in-gap) phase, resulting in only minor variations in the optical absorption edge under compression. Therefore, despite the pressure-induced structural transition from the low-pressure $C2/c$ phase to the high-pressure $P\bar{3}1c$ (in-layer) phase, the optical absorption edge remains largely preserved across the investigated pressure range.



\section{\label{sec:conclusion}Conclusions}

In this work, we investigated the pressure-dependent properties of LiInP$_2$S$_6$ polymorphs, including the monoclinic $C2/c$ and trigonal $P\bar{3}1c$ (in-layer and in-gap) phases, using first-principles DFT calculations. 
Our results reveal that hydrostatic pressure drives a phase transition from the ground-state monoclinic $C2/c$ phase to a trigonal $P\bar{3}1c$ in-layer phase at 0.38\,GPa pressure, which is mediated by enhanced interlayer coupling and anisotropic lattice compression. 
Despite this structural reorganization, all three phases remain mechanically stable under compression, while exhibiting distinct pressure-dependent mechanical and elastic responses. The elastic wave velocities and Debye temperatures consistently increase with pressure, reflecting progressive lattice stiffening and strengthening of bonding interactions. 
Remarkably, the electronic band gaps and optical absorption edges are largely insensitive to pressure within each phase, changing by only a few percent up to maximum 13\% at 10\,GPa; however, they undergo a pronounced discontinuity across the pressure-induced phase transition. 
These findings establish LiInP$_2$S$_6$ as a pressure-tunable ionic–vdW material in which subtle competition between interlayer interactions and ionic configurations governs structural stability, mechanical response, and electronic functionality, offering a broader framework for understanding pressure-driven phase engineering in layered quantum materials.


\section*{Acknowledgment}
Authors acknowledge support from the U.S.~Department of Energy, Office of Science, Office of Fusion Energy Sciences, Quantum Information Science program under Award No. DE-SC-0020340. Authors also thank the Pittsburgh Supercomputer Center (Bridges2) supported by the Advanced Cyberinfrastructure Coordination Ecosystem: Services \& Support (ACCESS) program, which is supported by National Science Foundation Grants No.~2138259, No.~2138286, No.~2138307, No.~2137603, and No.~2138296

\appendix

\section{Lattice parameter evolution under pressure}\label{sec:lattices}

Table~\ref{tab:lattice} presents the optimized ambient-pressure lattice parameters and unit-cell volumes calculated using the DFT-D3 method for the three studied phases: $C2/c$, $P\bar{3}1c$ (in-layer), and $P\bar{3}1c$ (in-gap).

\begin{table}[h]
\centering
\caption{\label{tab:lattice}Optimized lattice parameters and unit-cell volumes of the studied LiInP$_2$S$_6$ phases at ambient pressure}
\begin{tabular}{lccc}
\hline\hline
Structure & $a_0=b_0$ (\AA) & $c_0$ (\AA) & $V_0$ (\AA$^3$) \\
\hline
$C2/c$                        & 6.02 & 12.91 & 401.2 \\
$P\overline{3}1c$ (in-layer) & 6.02 & 12.66 & 397.6 \\
$P\overline{3}1c$ (in-gap)   & 6.05 & 12.73 & 403.5 \\
\hline\hline
\end{tabular}
\end{table}

\section{Dielectric function calculations}
\label{sec:dielectric}

\begin{figure}[htbp]
  \centering
  \includegraphics[width=\columnwidth]{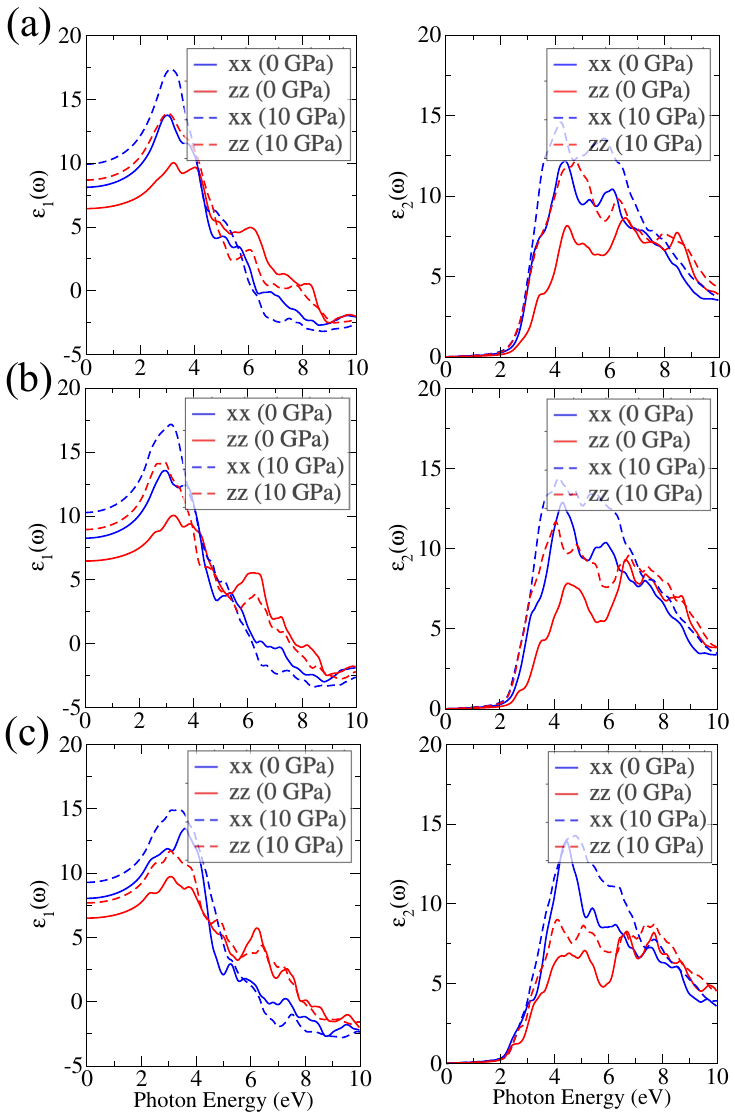}
  \caption{The real ($\varepsilon_{1}(\omega)$) and imaginary ($\varepsilon_{2}(\omega)$) 
  parts of the dielectric function for (a) $C2/c$, (b) $P\bar{3}1c$ (in-layer), and 
  (c) $P\bar{3}1c$ (in-gap) phases. For all phases, the dielectric tensor components 
  satisfy $\varepsilon_{xx} = \varepsilon_{yy}$. Solid lines correspond to ambient pressure, 
  while dashed lines represent the results at 10~GPa.}
  \label{fig:dielectric}
\end{figure}

This appendix presents the calculated dielectric functions of the three studied LiInP$_2$S$_6$ phases. Here, $\varepsilon_1(\omega)$ and $\varepsilon_2(\omega)$ correspond to the real and imaginary parts of the dielectric function, respectively. The calculated dielectric responses at ambient pressure and 10~GPa are shown in Fig.~\ref{fig:dielectric}.

\section{Optical Properties}
\label{sec:optical}
\begin{figure*}[t]
  \centering
  \includegraphics[width=0.75\textwidth]{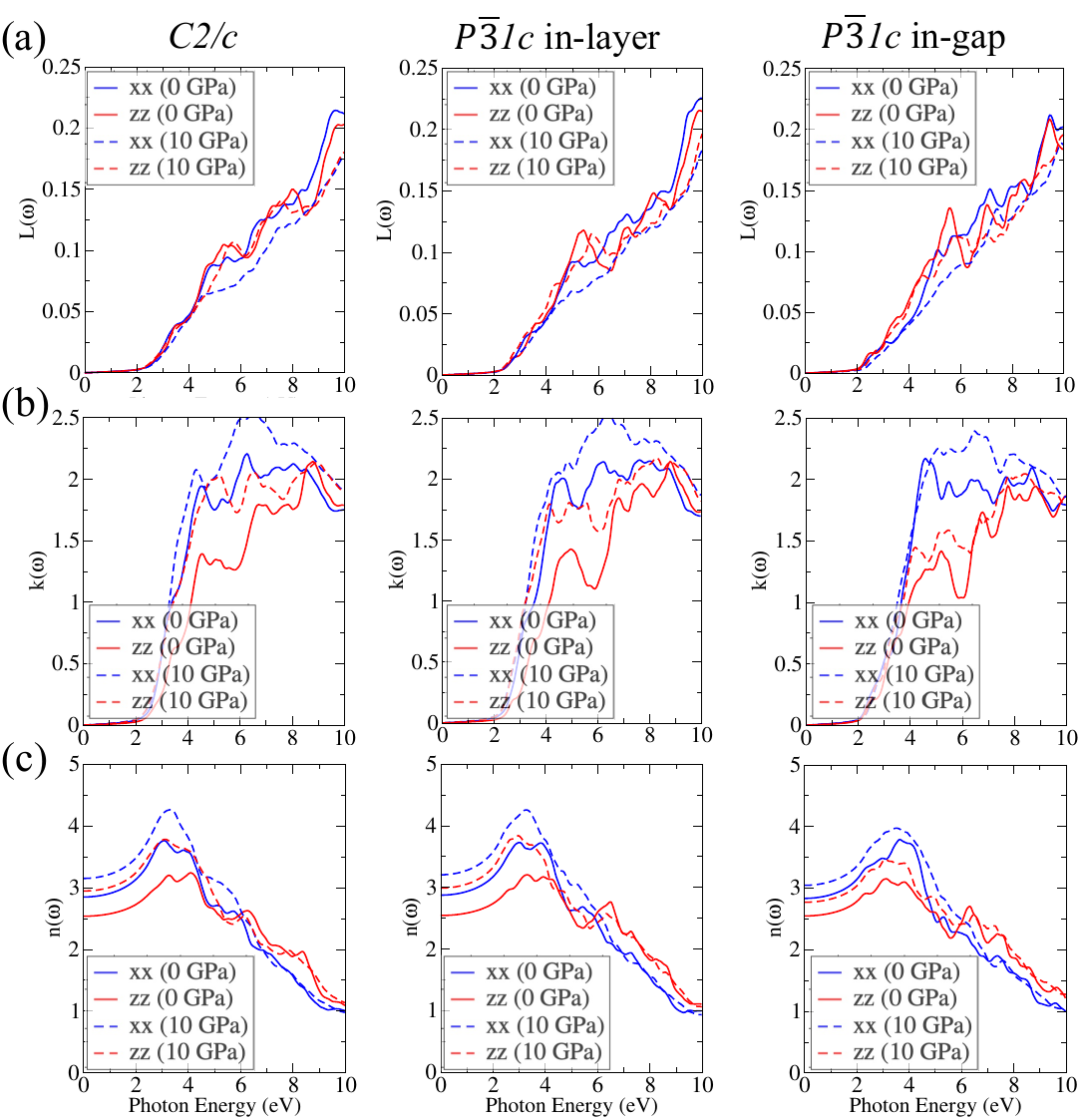}
  \caption{Calculated optical properties of LiInP$_2$S$_6$: (a) energy-loss function, (b) extinction coefficient, and (c) refractive index for the $C2/c$, $P\bar{3}1c$ (in-layer), and $P\bar{3}1c$ (in-gap) phases. For all phases, the tensor components satisfy $M_{xx} = M_{yy}$. Solid lines correspond to ambient pressure (0~GPa), while dashed lines represent results at 10~GPa.}
  \label{fig:optical_prop}
\end{figure*}

In this appendix, we present the evolution of the optical properties of the three studied phases under applied pressure, as shown in Fig.~\ref{fig:optical_prop}. The energy-loss function $L(\omega)$, shown in Fig.~\ref{fig:optical_prop}(a), is associated with plasmon excitations. Under applied pressure, the main peaks shift toward higher photon energies, indicating stronger interatomic bonding and higher excitation energies for electronic transitions. The extinction coefficient $k(\omega)$ [Fig.~\ref{fig:optical_prop}(b)] reflects the absorption strength of the material and exhibits slightly increased peak intensities under pressure, suggesting that the optical absorption remains relatively robust due to the preserved orbital hybridization under compression, in agreement with the DOS analysis. The refractive index $n(\omega)$, shown in Fig.~\ref{fig:optical_prop}(c), describes light propagation within the material and shows a slight increase under pressure for all phases, indicating an enhanced polarization response arising from stronger interatomic interactions. Overall, the optical properties demonstrate a robust preservation of the absorption edge and spectral features under pressure. Despite the pressure-induced structural phase transition, the optical response of the LiInP$_2$S$_6$ phases remains largely unchanged.

\clearpage

\bibliography{References}
\end{document}